\documentstyle[preprint,aps]{revtex}
\input {psfig}
\begin{document}

\draft

\title{Dilepton emission from an equilibrating \\ 
non-abelian plasma}

\author{Gouranga C Nayak \thanks{ Electronic address:gcn@iitk.ernet.in}}
\address{Department of Physics, Indian Institute of 
Technology, Kanpur -- 208 016, INDIA }

\maketitle

\begin{abstract}
We study dilepton production from a thermally equilibrating 
quark-gluon plasma expected to be 
formed in ultra relativistic heavy-ion collisions.
The pre-equilibrium dynamics of quark-gluon plasma is studied 
within the color flux-tube model by solving non-abelian relativistic
transport equations. This dilepton rate crucially depends on the
collision time of the plasma. The results are compared with the Drell-Yan
productions. We suggest that a measurement of this rate at RHIC and LHC
will determine the initial field energy density.

\end{abstract}

\vspace{2.0cm}


\vspace{1.5cm}
 

\vspace{1.0cm}

\newpage

One of the primary aims of the ultra relativistic heavy-ion
collision experiment is to detect a new phase of matter, ${\it viz}$
quark-gluon-plasma(QGP). It is known from lattice quantum chromodynamics 
that hadronic matter undergoes a transition to this quark-gluon plasma phase
at very high temperatures
($\sim 200 MeV$) and densities ($\sim 2 GeV/fm^3$) \cite{karsch}.
While such a phase did exist in the early universe,
after $\sim 10^{-4}$ seconds 
of the big bang, it is interesting if we can recreate the early
universe and experimentally verify this QCD phase transition in the 
laboratory {\it i.e.} in high energy heavy-ion collisions.
In the near future, the 
relativistic heavy ion collider(RHIC) at BNL (Au-Au collisions at 
$\sqrt{s}=200 GeV$) and 
large Hadron Collider(LHC) at CERN (Pb-Pb collisions at $\sqrt{s} \simeq 
6.4 TeV$) \cite{awes,gust}
will provide the best oppurtunity to study such a phase transition. 
As this plasma exist for a very short time ($\sim 10 fm$) 
and formed over a very small volume ($\sim 100 fm^3$), 
the direct detection of this phase is not possible in these experiments.
The proposed signatures for its detection
are therefore indirect, and the prominent ones are 
: 1) $J/{\psi}$ supression \cite{matsui},
2) electromagnetic probes such as dilepton and direct photon production
\cite{strickland,sinha}, and 3) strangeness enhancement
\cite{rafelski}.

It has been suggested by Matsui and Satz that $J/{\psi}$ 
suppression is a good probe \cite{matsui} for the detection of 
an equilibrated quark-gluon plasma. Their
calculation is based on lattice QCD which
assumes equilibration of the plasma. However, in heavy-ion
collisions such as at RHIC
and LHC the assumption of such an equilibration of quark-gluon plasma 
is only suggestive. One does
not know exactly when the plasma equilibrates and there is no concrete 
calculation on this aspects. This sets
an uncertainty in the calculation of $J/{\psi}$ survival probability 
which is determined from screening. In this regard, there are calculations of
$J/{\psi}$ suppressions in the equilibrating quark-gluon plasma
\cite{gcn,xu} using short-distance QCD, but 
the uncertainties in the $J/{\psi}$ formation time and quark-gluon plasma 
equilibration time make it difficult to compare the
results with the experiments (see Ref. \cite{gcn} for the details).
The suppression of $J/{\psi}$ infact, been observed in
reactions where there is no 
quark-gluon plasma phase, such as in p-A collisions and in collisions
of light nucleus \cite{na38}. This suppression is well
explained by the nuclear absorption of $J/{\psi}$
\cite{huffn}. 
Although, recently, NA50 collaboration \cite{na50} reports an
excess in the supression of $J/{\psi}$, it is still not clear 
if an equilibrated quark-gluon plasma has formed in this collisions.
There are proposals that the data are explained 
by a deconfined partonic medium \cite{satz}, or by
a medium with high density \cite{blazoit}; 
but there are also other calculations which explain the
data without assuming any quark-gluon plasma phase \cite{hwa,wong}.
As far as the $J/{\psi}$ supression 
in nucleus-nucleus collision is concerned, many aspects of it 
has to be studied in greater detail before unambigious conclusions
can be drawn about the existence of quark-gluon plasma.
It thus appears that there is a need to study more than one signature
if one has to detect QGP. With that in mind we investigate 
another signature, {$\it i.e.$} dilepton emission.

Dileptons and single photons have long been proposed as useful
probes of the plasma \cite{sinha}, as once produced,
they hardly interact with the strong matter 
and thus carry the details of the circumstances
of their production. Being electromagnetic in nature, they 
do not suffer from
the final state ineractions (interactions with hadrons)
and hence keep the memory of their 
formation surroundings (such as the information about the 
temperature and number density). The major processes for dilepton 
production in ultrarelativistic heavy-ion collisions (URHIC)
are, (i) hard parton scattering, (ii) electromegnetic 
decay of hadrons and (iii) production from partons present in
QGP (thermal production).
The hard parton scatterings produce high $p_t$ lepton pairs 
which are calculated from pQCD, knowing the structure functions
of partons inside nucleus.
On the other hand, electromagnetic decay
of produced hadrons is the main source of dilepton and photon
production in heavy-ion collisions which
obscure the signal of interest (thermal production),
which are produced from QGP. 
The thermal emission of high invariant mass 
dilepton is calculated in the same way as that of
hard parton scatterings by using the thermal distributions of partons inside
QGP. It is expected 
that in high energy heavy-ion collisions such as at RHIC and LHC,
the thermal production of dileptons 
will be more than that from other processes \cite{dks}.
On the experimental side, 
an enhancement of dilepton yield is observed in central
200A GeV S+Au \cite{gr1}, S+W \cite{gr2},
and 160A GeV Pb+Au \cite{gr3} collisions.
However, these data are partially explained by the conventional
mechanism of binary hadron
collisions, {\it e.g.}, by $\pi \pi \rightarrow e^+ e^-$ 
processes \cite{gr4,gr5,gr6} and 
by the contribution of the collective bremsstrahlung mechanism \cite{gr7}.
In ultrarelativistic heavy-ion collisions
such as at RHIC and LHC situation might be different and one may expect an
enhancement of the dilepton production 
which can be due to the formation of quark-gluon plasma.

For a thermally equilibrated quark-gluon plasma, 
the distribution of dilepton is given by \cite{dks}
\begin{equation}
{dN \over {dy dM^2}} = \pi R^2 {3 M^2 \sigma(M) T_0^2 \tau_0^2 \over 
{2(2\pi)^4}} {T_0^4 \over {M^4}} \times [H(M/T_0)-H(M/T_c)]
\end{equation}
where $T_c$ is the freeze-out temperature at which one stops the hydrodynamic
evolution of the quark-gluon plasma, or perhaps the critical temperature if 
there is a phase transition. Here M is the invariant mass of the dilepton pair, 
R is the size of the nucleus and $\sigma$ is the anihilation cross section
for quark-antiquark into a dilepton pair. The function H is written in terms
of Bessel function $K_1$ as, $H(z) = z^2(z^2+8)K_0(z)+4z(z^2+4)K_1(z)$.
Assuming a fast thermalisation and high initial temperature, 
enhancement in the dilepton rate was 
found by several authors \cite{dks}.
However, it can be seen from the above equation that there is always
an uncertainty in the dilepton rate because of the assumption of initial
temperature ($T_0$) and initial time ($\tau_0$), at which quark-gluon plasma
thermalise. The dilepton rate in any realistic calculation depends crucially
on these initial conditions which determines the hydrodynamic evolution.
For this reason
a detailed calculation of the plasma evolution in different stages 
of ultrarelativistic heavy-ion collisions is necessary.
The various stages by which the evolution of 
quark-gluon plasma is described in URHIC are, i) 
pre-equilibrium, ii) equilibrium, where one actually studies
the equilibrated quark-gluon plasma, 
and iii) hadronisation. The pre-equilibrium stage
of the collision which leads to thermal and then chemical equilibrium
has a crucial role to play in the equilibration of the plasma and also on the
calculation of different signatures.
>From this point of view it is necessary to study what happens 
to the dilepton production from different stages of QGP,
rather than estimating it in an equilibrated quark-gluon plasma (Eqn.-1 ).
In this paper we study the dilepton rate for a thermally equilibrating
quark-gluon plasma in ultra relativistic heavy-ion collisions.

The rate of equilibration of quark-gluon plasma in URHIC 
is different for diiferent models.
One of the relevant model that describes the production and the
equilibration of QGP in URHIC 
is the color flux-tube model \cite{baym,kajantie,banerjee}.
This model is a generalisation of the familiar Lund string model
widely used for $e^{+}e^{-}$ and $p-p$ collisions \cite{andersson}.
Within this model, two nuclei that undergo a central collision at
ultra high energies are highly lorentz-contracted as thin plates. 
When these two highly lorentz-contracted nuclei pass through each other
they  acquire a nonzero color charge ($<Q> = 0, <Q^2> \neq 0$), by 
a random exchange of soft gluons. The nuclei which act as color 
capacitor plates produce a chromo-electric field between them
\cite{low,nussinov}. 
This strong electric field creates $q\bar{q}$ and gluon pairs
via the Schwinger mechanism \cite{schwinger}
which enforces the instability of the
vacuum in the presence of an external field. The partons so
produced, collide with each other and also get accelerated
by the background field. In the case at hand, the color degree has a central
dynamical role in the evolution of the plasma.
In our recent studies we have incorporated this dynamics in the plasma
evolution in heavy-ion collisions \cite{nayak,nayak1}.
The color charge($Q^a$) which is a vector in the color space
obeys Wong's equation \cite{Wong}:
\begin{equation}
{dQ^a\over d{\tau}}= f^{abc} u_\mu Q^b A^{c\mu}. 
\end{equation}
where $A^{a\mu}$ is the gauge potential, and
$f^{abc}$ is the structure constant of the gauge group.
This equation describes the precission of the color charge in external
chromo field.
The non-abelian version of the Lorenz force equation 
\begin{equation}
{dp^{\mu} \over d{\tau}} = Q^a F^{a\mu\nu}u_{\nu},
\end{equation}
describes the acceleration of the color particles by the background chromo
field $F^{a\mu\nu}$.
Due to these dynamical nature of the color charge,
the usual classical phase space 
of coordinate and momenta is extended to include color.
The single particle distribution function $f(x,p,Q)$ of quark and gluon is 
then defined in the compact phase space of dimension 14 in SU(3).
In this extended
phase space a typical relativistic transport equation is written as \cite{heinz}
\begin{equation}
\left[ p_{\mu} \partial^\mu + Q^a F_{\mu\nu}^a p^\nu 
\partial^\mu_p
+ f^{abc} Q^a A^b_\mu
p^{\mu} \partial_Q^c \right]  f(x,p,Q)=C(x,p,Q)+S(x,p,Q).
\end{equation}
One needs to solve this transport equation to study the non-equilibrium
dynamics of the quark-gluon plasma expected to be formed in ultra relativistic
heavy-ion collisions.
The first term in the above non-abelian relativistic transport equation
corresponds
to the usual convective flow (free streaming expansion), the 
second term is the non-Abelian version of
the Lorentz force term and the 
last term corresponds to the precession
of the color charge as described by Wong's equation.
$S$ on the right hand side of equation (4) 
correspond to the source term for parton production from background
chromoelectric field via Schwinger non-perturbative mechanism. $C$ corresponds
to the collision term. In general we have to write
separate equations for quarks, antiquarks and gluons since they
belong to different representations of the gauge group.
For anti-quarks the distribution function $\bar{f}(x,p,Q)$ obeys
a simillar equation, with $Q^a$ replaced by $-Q^a$ ({\it i.e.} the
second term in the above equation changes sign). These equations, which
are lorentz and gauge invariant \cite{heinz}, 
are closed with the Yang-Mills equation,
\begin{equation}
(D_{\mu}F^{\mu \nu})^a(x) = j^{\nu}(x)=g\int p^{\nu} Q^a [f_q(x,p,Q)
-\bar{f}_q(x,p,Q) + f_g(x,p,Q)]dP dQ.
\end{equation}
to study the production and equilibration of quark-gluon
plasma.

According to Bjorken's proposal \cite{bjorken} the distribution function
and other physical quantities are
written in terms of the boost invariant parameters
$\tau ( = \sqrt{t^2 - z^2})$, $p_t$ and $\xi ( = \eta - y)$.
Here $\eta$ is the space time rapidity ($tanh\eta = z/t$) and
$y$ is the momentum rapidity ($tanh{y} = p_z/p^0$).
In our calculation we have employed a collision term 
in the relaxation time approximation: 
\begin{equation}
C = {-{p^\mu u_\mu (f-f^{eq})} \over {\tau_c}},
\end{equation}
where $f^{eq}$ is the local eqilibrium distribution function with 
explicit color dependences. This is given by \cite{heinz3}: 
\begin{equation}
f_{q,g}^{eq} = \frac{2}{\exp ((p^\mu-Q^a \cdot A^{\mu a}) 
u_\mu /T(\tau)) \pm 1}.
\end{equation}
The +(-) sign in the above expression is for quarks(gluons).
The source term $S$ for $q\bar{q}$ and gluon pair production
is obtained by the Schwinger mechanism of particle production
\cite{nayak1}.
In the process where field and particles are present, we use the 
energy momentum conservation equation: 
\begin{equation}
\partial_\mu T^{\mu\nu}_{matter} + 
\partial_\mu T^{\mu\nu}_{field}=0,
\end{equation}
with $T^{\mu\nu}_{matter}=\int p^\mu p^\nu (2 f_q + 2 \bar{f}_q
+ f_g)d\Gamma dQ$
and
$T^{\mu\nu}_{field} = \mbox{diag} (E^2/2 ,E^2/2 ,E^2/2 ,-E^2/2 )$.
Here $d \Gamma = d^3 p/{(2 \pi)^3 p_0} = p_t d p_t d \xi/{(2 \pi)^2}$,
and $dQ$ is the integral in the color space.
The factor 2 in the above expression is for two flavors of massless quark. 
Equation (4) along with equation (8) is solved numerically by a double
self consistent method \cite{fnote1} 
to study the equilibration of
quark-gluon plasma in URHIC.
In the earlier paper \cite{nayak1} we have studied the bulk properties 
and the equilibration of quark-gluon plasma (which occurs arond 1 fm).
Here we calculate the dilepton production rate using these distribution
functions of partons $f(\tau, \xi, p_t, Q)$.

The dominent process which produces a dilepton pair $l^+ l^-$ is 
\begin{equation}
q + \bar{q} \rightarrow \gamma^{*} \rightarrow l^+ + l^-,
\end{equation}
where $\gamma^{*}$ is the intermediate virtual photon.
The dilepton emission rate $dN$ for such a process
in a space time volume $d^4x$ is 
\begin{equation}
{dN \over {d^4x d^4P}} = {1 \over {(2 \pi)^6}} \int d^3p_1 d^3p_2 dQ f(x, p_1, Q) \bar{f}(x, p_2, Q) v_{rel} \sigma(M^2) \delta^4(P- p_1 - p_2).
\end{equation}
Here $P^{\mu}$ is the four momentum of the lepton
pair, ${P_T}$ is the transverse momentum, 
$M_T (= \sqrt{M^2 + {\bf P_T}^2})$ is its transverse mass and 
M is the invariant mass ($M^2 = P^{\mu} P_{\mu}$).
$v_{rel}( = M^2/{2 E_1 E_2})$ is the relative flux velocity of quark and
antiquark pair in the above process.
The dilepton production cross section $\sigma(M^2)$ for the above reaction is 
\begin{equation}
\sigma(M^2) = {{4 \pi \alpha^2} \over {3 M^2}}
[1+{ 2 m_l^2 \over{M^2}}] [1-{4 m_l^2 \over {M^2}}] F_q,
\end{equation}
with $F_q = N_s^2 {1 \over {N}} \sum_f e_f^2$ and $m_l$ is the mass of
the lepton.
Here, N is the color averaging factor (which corresponds to the
volume of the color space),
$N_s$ is the spin degeneracy ($N_s=2 s + 1$)
and $e_f$ is the fractional charge of the flavour.
For dilepton pair of large invariant mass 
the emission rate in the midrapidity region
(Y = 0) is given by:
\begin{equation}
{dN \over {dM_T^2 dY dP_T^2}} = {{5 R^2 \alpha^2}
\over {72 \pi^7}} \int d\tau \tau W(f_1, f_2). 
\end{equation}
where
\begin{equation}
W(f_1, f_2) = \int^{+ \infty}_{- \infty} d\eta \int^{+ 
\infty}_{- \infty} d\xi_1 \int^{p_+}_{p_-} dp_{t1} 
\int dQ {p_{t1} f(\tau, p_{t1}, \xi_1, Q) \bar{f}
(\tau, p_{t2}, \xi_2, Q) \over {[p_{t1}^2 P_T^2 -[p_{t1}
M_t ch(\eta - \xi_1) - {1 \over {2}} M^2]^2]^{1/2}}} \nonumber
\end{equation}
with
$p_{t2} = \sqrt{M_T^2 - 2 M_T p_{t1} ch(\eta -
\xi_1) + p_{t1}^2}, \hspace{0.2cm} sh{\xi_2} = 
{1 \over {p_{t2}}} (M_T sh\eta - p_{t1} sh\xi_1)$ 
and $p_{\pm} = {1 \over 2} M^2 [M_T ch(\eta - \xi_1) \mp P_T]^{-1}$.
In equation (12) we have used
$d^4x = \pi R^2 d\tau \tau d\eta$ (where R is the radius of the
nucleus). $\alpha (=1/137)$ is the coupling constant of the
electromagnetic interaction.

It can be mentioned here that the thermal 
dilepton rate does not depend on $M$ and
$P_T$ separately, but only on $M_T$ \cite{mclerran}. 
However this $M_T$ scaling is
violated for an equilibrating plasma, which is seen in equation (12).
In this case 
the dilepton rate depends on both $M_T$ and $P_T$.
This would also be the case if one takes the transverse expansion of the
plasma into account \cite{mcl}. 

We will now present our results of the dilepton rate using the above
non-equilibrium distribution function, $f(\tau, \xi, p_t, Q)$,
of quarks and antiquarks. In our calculation we have taken 
$R= 7 fm$ and the initial field energy
density $\epsilon_0 (= 1/2 E_0^2)$ to be equal 
to 300 GeV/$fm^3$ (see Ref. \cite{nayak1}).
We have considered two different cases corresponding to two different 
relaxation times, $\tau_c = 5 fm$ and $\tau_c = 0.2 fm$.
The relaxation time $\tau_c = 5 fm$
corresponds to collisionless limit and
$\tau_c = 0.2 fm$ corresponds to a more realistic 
limit of equilibration \cite{geiger1}.

We have calculated the dilepton rate as a 
function of $M_T$ for $P_T = 0 $
and $1$ GeV respectively. 
In Fig-1 we have presented the results for 
$\tau_c =0.2 fm$.  
As can be seen from the figure, the dilepton yield becomes 
smaller for higher values of transverse momenta.
In Fig-2 we have presented our results
for $\tau_c = 5 fm$ which corresponds to the collisionless limit.
In this case the rate at higher $M_T$ is found to be larger 
than that at $\tau_c = 0.2 fm$. This 
is because the average energy per parton is higher at $\tau_c = 5 fm$
than at $\tau_c = 0.2 fm$ as observed earlier \cite{nayak1}.
The average energy per parton is around 4 GeV for $\tau_c$ = 5 fm and
is around 2 GeV for $\tau_c$ = 0.2 fm.
In the absence of any collision the partons come with higher energies
than the partons with collision. From this point of view it is crucial
to determine the collision time accurately. We will 
mention here that a determination of the collision time $\tau_c$ is possible
by a triple self consistent numerical method \cite{fnote2} instead of a
double self consistent method (which we have employed here to solve the
transport equations).

In order to get a feeling, how crucial the dilepton production from 
an equlibrating QGP is, we have compared our results with the Drell-Yan
productions. The Drell-Yan results are taken from the Ref. \cite{asakawa1}.
These spectra are calculated at pp center-of-mass energy 
$\sqrt{s}$ = 200 GeV scaled to UU central collisions assuming a simple
factorisation of the nuclear mass dependences. 
It can be seen in Fig-3 that 
the dilepton rate from the pre-equilibrium stage is larger than
the Drell-Yan production
for very small transverse momentum ($P_T \simeq 0 GeV$).
This is true when $M_T \le 2 GeV$. For $M_T \ge 2 GeV$,
the Drell-Yan production dominates over the
pre-equilibrium dilepton production.
For large transverse momentum ($P_T$ = 1 GeV) of 
dilepton pair, the production from the pre-equilibrium stage
is smaller than the Drell-Yan production in the whole range
of dilepton transverse mass. This has been shown in Fig-4.
This result is contrary to the earlier findings \cite{bialass,geiger2}
where it is shown that dilepton production from the pre-equilibrium
stage dominates over the Drell-Yan production.
The partons formed are not too
hard in the color flux-tube model, and one expects to have 
such a low rate in the dilepton spectra than the Drell-Yan emissions, 
which are produced from the primary hard scattering of partons.
However, the comparisons are not very strict because the choice of 
initial field energy density $\epsilon_0$ (= 1/2 $E_0^2$, $E_0$ being the
initial chromoelectric field) is arbitary and there is no way of determing
this. This is because one does not know how many soft gluons are exchanged
(which deterimines the strength of the initial chromoelectric field) 
when two nuclei cross
each other in ultra relativistic heavy-ion collisions. 
This initial condition can be determined from the measurement of some 
experimental signatures, such as dilepton production.
However, the results we have found here do not seem unnatural.
This is because, if one dentifies the energy deposited at RHIC \cite{bjorken}
(when two nuclei are in maximum overlap) with  
the initial field energy, the initial field 
energy density becomes $\sim$ 250 GeV/$fm^3$ \cite{geiger1}. 
In any case, we hope to extract the inital field energy density from the
measurement of dilepton spectra at RHIC and LHC (this is done
only after a careful substraction of the dilepton rates from other
processes in different stages of quark-gluon plasma). After the determination of
this initial field energy density, other bulk properties of the plasma
can be determined accurately.

Summarizing the paper, we have calculated the dilepton spectra
in ultra relativistic heavy-ion collisions within color flux-tube
model, with non-abelian features explicitly incorporated.
The production is larger in the
collisionless limit of the plasma. After one determines the collision time
accurately by a triple self consistent procedure 
\cite{fnote2}, an experimental measurement of the 
dilepton spectra at RHIC and LHC will shed light on the determination of
initial field energy density which we have assumed in our model.
Only after this, the predictons of all the bulk properties of the plasma
such as temperature and number density, will be determined accurately.

\vspace{2.0cm}

\subsection*{Figure captions}

\noindent
{\bf FIG.~1.} Dilepton rate from pre-equilibrium stage,
as a function of $M_T$ for $\tau_c=$0.2 fm.

\noindent
{\bf FIG.~1.} Dilepton rate from pre-equilibrium stage,
as a function of $M_T$ for $\tau_c=$5.0 fm (collisionless limit).

\noindent
{\bf FIG.~3.} Drell-Yan and pre-equilibrium dilepton rate as
a function of $M_T$ for $\tau_c$=0.2 fm ($P_T=0$ GeV).

\noindent
{\bf FIG.~4.} Drell-Yan and pre-equilibrium dilepton rate as
a function of $M_T$ for $\tau_c$=0.2 fm ($P_T=1$ GeV).

\begin{figure}[tb]
\vspace*{0.3cm}
\centerline{\hspace*{1cm}
\psfig{width=15.2cm,figure=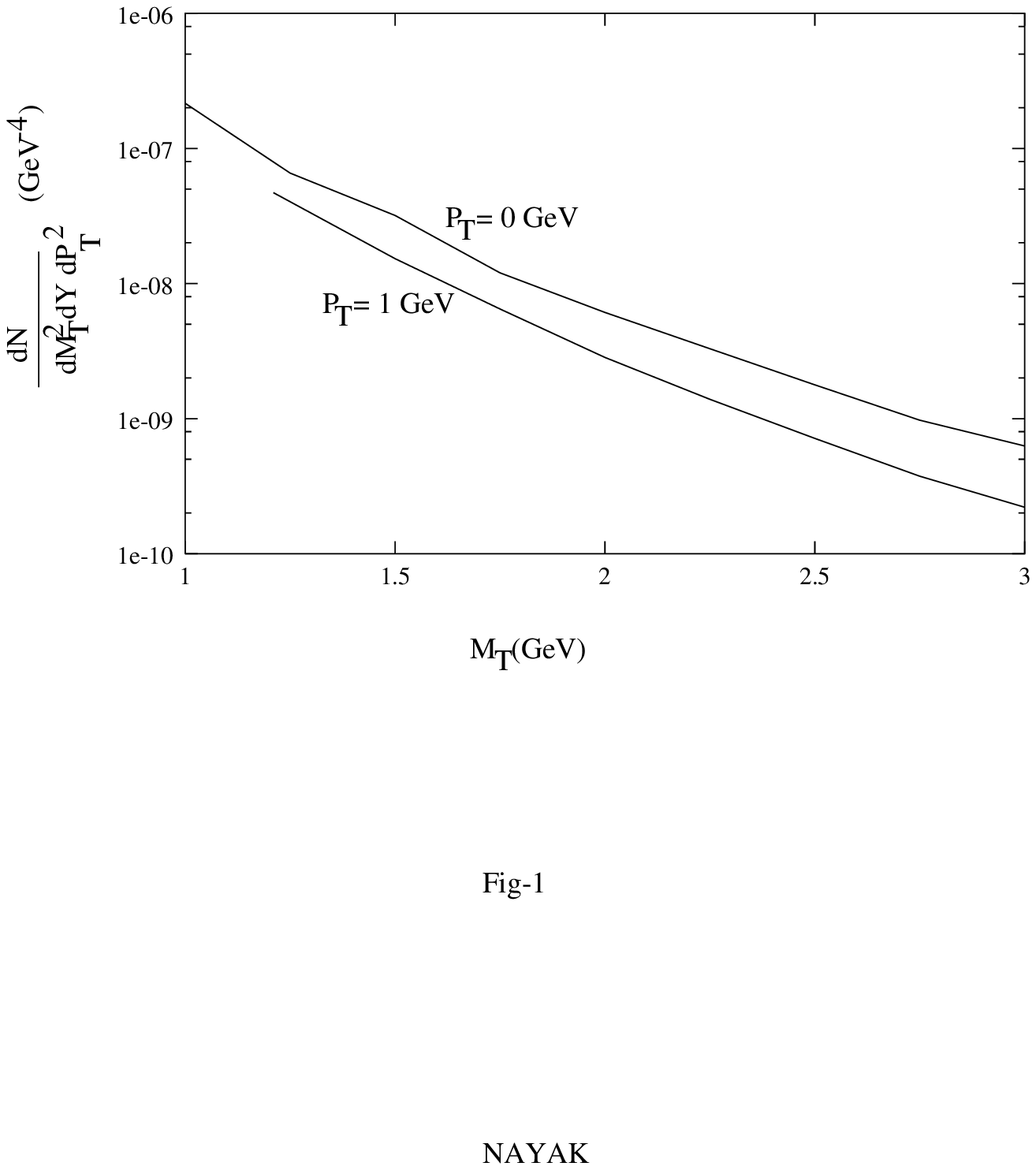}}
\vspace*{-2.cm}
\end{figure}

\begin{figure}[tb]
\vspace*{0.3cm}
\centerline{\hspace*{1cm}
\psfig{width=15.2cm,figure=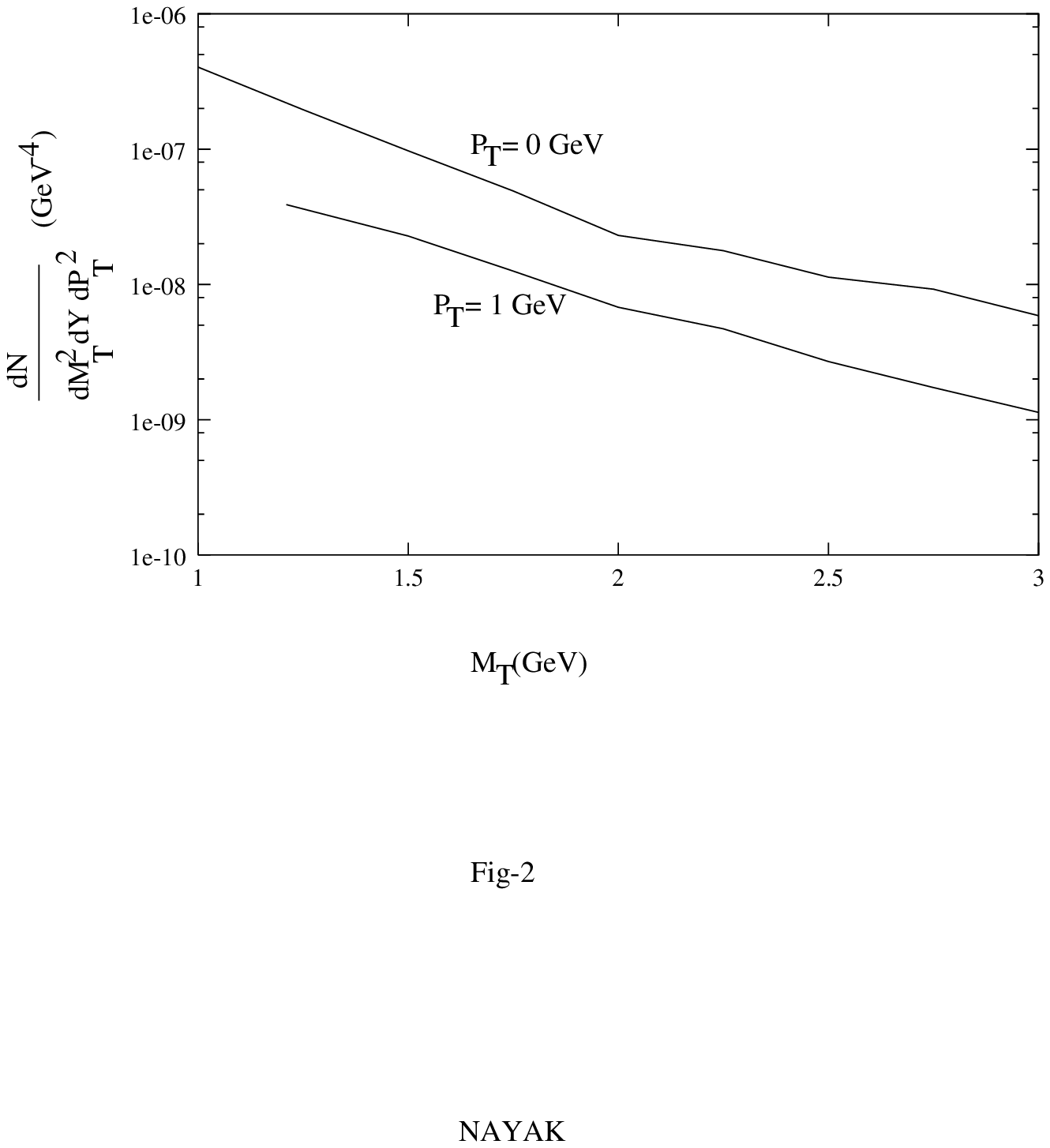}}
\vspace*{-2.cm}
\end{figure}

\begin{figure}[tb]
\vspace*{0.3cm}
\centerline{\hspace*{1cm}
\psfig{width=15.2cm,figure=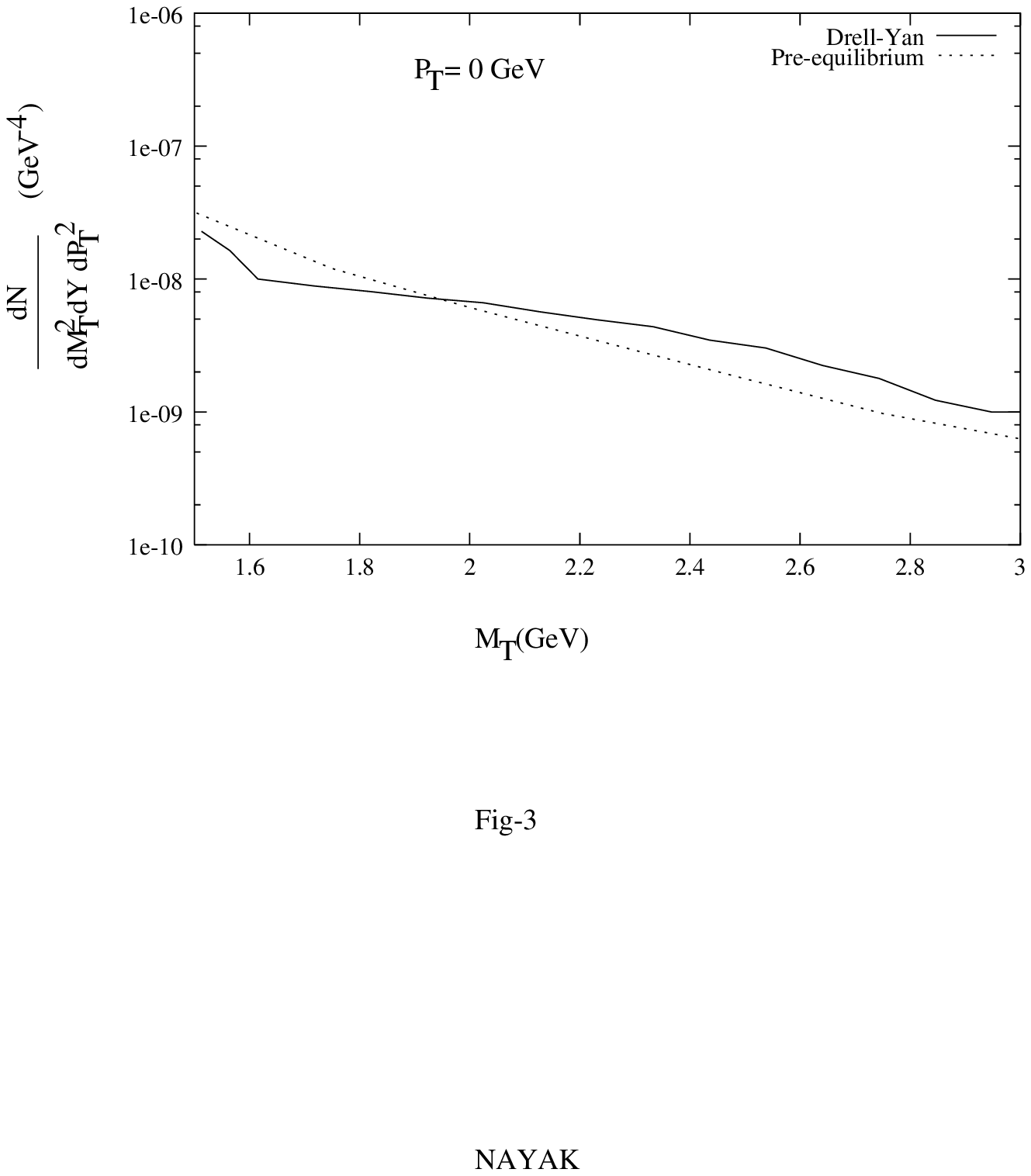}}
\vspace*{-2.cm}
\end{figure}

\begin{figure}[tb]
\vspace*{0.3cm}
\centerline{\hspace*{1cm}
\psfig{width=15.2cm,figure=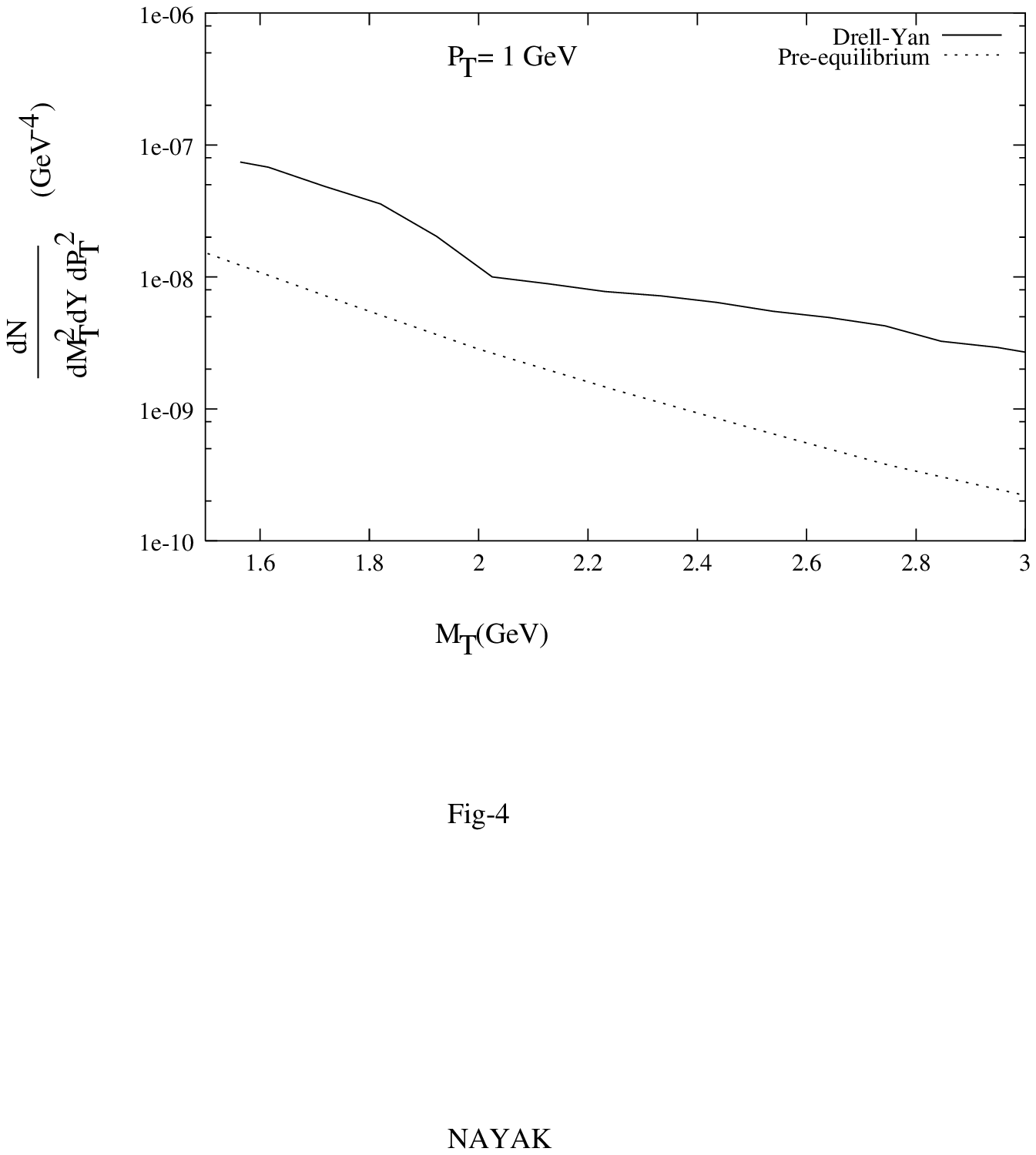}}
\vspace*{-2.cm}
\end{figure}


\begin{thebibliography}{99}
\bibitem{fnote1} The numerical program is described in Ref. \cite{nayak,nayak1},
       where the determination of E and T is done by a double self consistent method.
       In this calculation an equilibration is assumed to relate temperature T to the
       particle energy density.
\bibitem{fnote2} The relaxation time $\tau_c$ is determined in 
       simillar way (see Ref. \cite{fnote1}) by relating $\tau_c$ to the
       number density of the plasma, assuming equilibration.
\end{thebibliography}
\end{document}